# LUMINESCENCE OF NATURAL α-QUARTZ CRYSTAL WITH ALUMINUM, ALKALI AND NOBLE IONS IMPURITIES

A.N. Trukhin[1]

Institute of Solid State Physics, University of Latvia, Latvia

ABSTRACT

Photoluminescence and thermally stimulated luminescence of synthetic and natural (morion and smoky) α-quartz crystals doped with aluminum and alkali ions were studied. The samples were examined both untreated and subjected to substitution of alkali ions for copper or silver ions. The photoluminescence spectrum of the untreated crystals is characterized with the main blue band around 400 nm (~ 3.1 eV). The corresponding luminescence center is based on a defect containing aluminum and alkali as compensators in natural and synthetic quartz crystals. Photoluminescence is subjected to thermal quenching and can be detected at high temperatures above 700 K, however the main intensity decay takes place at 200 K. The thermal quenching activation energy is 0.15±0.05 eV and the frequency factor is $3 \cdot 10^{7\pm3}$ s$^{-1}$. In the samples with silver ions the main luminescence band is located at ~ 260 nm (~ 4.7 eV) with a time constant of ~ 37 μs at 80 K, and in the samples with copper ions the PL band is at ~ 360 nm (~ 3.4 eV) with a time constant ~ 50 μs at 80 K. The initial luminescence of crystals is greatly reduced after introduction of noble ions. The luminescence of noble ions quenches at 700 K without drop in intensity at 200 K. For luminescence associated with silver the energy of thermal quenching is 0.7±0.1 eV with a frequency coefficient of $1 \cdot 10^{14\pm3}$ s$^{-1}$, and for the luminescence related to copper, these parameters are 0.55±0.1 eV and $1 \cdot 10^{14\pm3}$ s$^{-1}$. The differences in intra-center luminescence properties of the same defect containing alkali ions or noble ions are based on differences in electronic transitions. In the case of alkali ions the charge transfer transitions between oxygen and alkali ions. In the case of noble ions absorption – luminescence corresponds to intra ion transitions. Radiation properties are related to trapping of an electron on one valence ion. Created atom moves out of aluminum containing defect. The hole remains on aluminum-oxygen defect. Thermally stimulated luminescence is related to release of atom, it diffusion to aluminum defect with the hole on oxygen and following radiative recombination. Optically stimulated luminescence is explained by the similar process of optical release of excited atom and movement to aluminum defect and recombination of electron with hole.

---

[1] Corresponding author.
E-mail address: truhins@cfi.lu.lv

INTRODUCTION

Many samples of natural crystalline α-quartz at low temperatures (77 K) exhibit the intense blue photoluminescence excited in the range of 6 eV [1]. Recent studies [2, 3] show that this luminescence appears in aluminum-containing quartz and is attributed to a certain complex of aluminum and alkali ions. The luminescence decay kinetics is exponential and the decay time constant is ~2 ms at 77 K [2,3]. It was noted that at liquid helium temperatures the blue photoluminescence decay curves of natural quartz change. Components appear one faster than 2 ms (around 0.6 ms) and one slower (around 12 ms) [2, 3]. The same effect was observed also in the luminescence of a self-trapped exciton (STE) in the pure samples of crystalline quartz [2]. This effect is explained by the splitting of the triplet state of the excited center in zero magnetic field, this is supported by observation of optically detected magnetic resonance (see references in [2]). Such similarity in the behavior of the kinetics at the temperature of liquid helium suggests the similarity of the electronic states of the excited center containing aluminum and alkali ions and the STE, [2].

Substitution of alkali ions for noble ions, performed in [4] for the case of copper and [5] for the introduction of silver in quartz, supports Al, $O^-$, $Me^+$ ($Me^+$ alkali or noble ion) model of the luminescence center. After the substitution, the initial luminescence of natural quartz disappeared and a new luminescence of the luminescence centers of copper and silver appeared. The similar situation is observed in silica glass [6,7]. A sample of silica with only alkali ions provides a luminescence similar to that known for pure alkali-silicate glasses [8]. Thermal quenching of PL of the original quartz crystal [2, 3] is observed in the temperature range above 200 K similarly to the STE luminescence, however, photoluminescence is also observed at room temperature. Luminescence centers related to alkali ions and noble ions have not studied yet at higher temperatures.

The X-ray excited luminescence of silver doped natural quartz was studied in [9]. The basic research task is to study the intra-center and recombination luminescence of a defect Al, $O^-$, $Me^+$. This research is important because natural quartz is often used in radiation measurements. The method is based on spectral and kinetics measurements at different temperatures using for the fundamental relation between life time (τ) of luminescence center in excited state and luminescence quantum yield (η) in intra-center processes [10]. In [10] the life time is determined as $1/\tau = p_r + p_{nr}$, where $p_r$ is probability of radiative transitions and $p_{nr}$ is probability of non-radiative transitions mainly due to thermal quenching. Quantum yield is determined as $\eta = p_r (p_r + p_{nr})^{-1}$. Time constant of non-quenched radiative transitions is $1/\tau_o = p_r$. Then $\tau(T)/\tau_o = \eta(T)/\eta_o$ is relation connecting life time and quantum yield, [10].

Ionizing radiation provides the induced absorption and the subsequent thermally and optically stimulated luminescence.

The task of the given work is to understand why the defects of natural quartz provide luminescence at high temperature and to explain the role of those defects in recombination processes.

EXPERIMENTAL

Samples of the study were natural crystals of quartz: morion and smoky. Both samples were studied as obtained, as well as annealed, to remove the coloration. Other samples were synthetic crystals with the addition of aluminum at a level of $10^{-1}$ % by weight. It is assumed that the alkali ions in the latter sample were introduced unintendedly due to the hydrothermal crystal growth method using alkali metal hydroxide.

The ArF laser (193 nm), model PSX-100, made by Neweks, Estonia, with a pulse energy of about 5 mJ and duration of 5 ns, was used to excite luminescence. For measurements of excitation spectra of photoluminescence, a 0.5 m Seja-Namioka vacuum monochromator with a deuterium lamp or spark discharge in hydrogen was used. The luminescence detection was carried out using a monochromator with a grating (MCD2) having both slits about 1 mm as well as Hamamatsu mini spectrometer model C10082CAH. Photomultiplier (H6780-04) with a 50 Ohm resistive load, and a photon counting module (H8959-01) have been employed. The digital oscilloscope (Tektronic TDS 2022B) was used to record the decay kinetics curves. Each curve was averaged for 128 pulses. Some measurements were made with a self-made multichannel analyzer. Time-resolved spectra were measured by recording the decay curve for each point of the PL spectrum in two time scales, one in the ns range of the other in the μs range. Time constant determination was made by the least square fit method in the semi log scale. Decay kinetics curve integration in time by the least square fit provides the time resolved intensities. Intensity of luminescence gives information about relative quantum yield. Noise of measured curves determines precision. The main method of measurement error determination is multiple observations in different measurements on different samples. A cryostat was used for low temperatures, and a resistor heater was used for high temperatures. Measurements at low temperature (above 15 K) were performed using the refrigerator and few measurements were made in cryostat with liquid helium (above 3 K achieved by pumping of helium). The temperature was measured with a resistor for low temperatures and a Cromel-alumel thermocouple for high temperatures. The irradiation was carried out using an X-ray tube with a tungsten anode in a 40 kV 20 mA mode. A special sample holder was applied, which makes it possible to cool up to 80 K and heat up to 750 K.

RESULTS

*Photoluminescence*

Fig. 1 shows the luminescence properties of natural and synthetic crystals at 80 K. There is a blue luminescence band at 3 eV, which positions slightly changes in different samples. Decay kinetics curve with time constant about 1.84 ms at 80 K is shown in Fig.1 inset. Also spectrum of time constant is shown with close squares. Time constant deviation within PL band probably is related to multiple types of centers. The PL and PLE spectra presented in the Fig.1 are normalized for easier comparison. These data are in good agreement with the previous study of natural quartz luminescence [2, 3]. It is remarkable that for synthetic quartz doped with aluminum with a high concentration, the observed luminescence is similar to that of natural quartz, Fig.1. Earlier [3], in synthetic quartz with a lower aluminum concentration, such luminescence was not detected. In Fig.2 the normalized PL spectra of natural quartz taken at different temperatures are presented by broken lines. It is seen that the luminescence is the same even at high temperatures and the maximum of the band is situated at about 3.1 eV. It has only having some deviation for different crystals.

Quartz samples exposed to high-temperature electrolysis in copper or silver electrodes show a greatly reduced initial luminescence, and besides, new luminescence bands assigned to $Cu^+$ or $Ag^+$, (Fig.2, lines). The maximum of the band for copper center is situated at 360 nm (3.45 eV) and that of silver center is at 260 nm (4.75 eV). If the noble ions activated samples subject to high-temperature electrolysis, using alkali halide crystals as electrodes (NaCl, KCl or LiF, for example), then luminescence due to noble ions reduced and the initial luminescence is renewed.

The temperature dependences of the PL intensity and the decay time constant are shown in Fig.3. The part (a) of Fig.3 labeled as Al – $Me^+$ provides dependences in samples not subjected to electrolysis with noble metals. Luminescence is detected even at 700 K. For these dependences three regions could be distinguished. 1) At low temperatures (3 - 20 K), intensity is changing little, whereas in decay kinetics curves two components of different decay rate appear. Above ~30 K both components merge together. 2) (150-200K). In this part of thermal dependence there is good correlation between behavior of luminescence intensity and decay time constant. Both are decreasing. 3) 200 -700 K. There is some difference between the temperature dependences of the PL intensity and the decay time constant for regions 2) and 3). The decay time constant is accelerated with increasing temperature, while the decrease in photoluminescence intensity is not analogous to τ(T). This can be explained by the influence of two factors on the electronic transitions in the center of luminescence. One of them is thermal quenching and another is an increase in the probability of the radiative transition with increasing temperature.

Thermal dependencies are expressed by Mott equations for thermal quenching [10]:

$I=I_0/(1+\tau_0 f \exp(-E/kT))$ (1)

$\tau=\tau_0/(1+\tau_0 f \exp(-E/kT))$ (2)

where I is intensity, $I_0$ – non-quenched intensity, τ – time constant, $\tau_0$ – non-quenched time constant, E activation energy and f is frequency factor. For Al – $Me^+$ center the obtained parameters are: E(I)=0.18±0.05 eV; E(τ)=0.2±0.5 eV; f(I)=3· $10^{7\pm3}$ $s^{-1}$; f(τ)~2.4· $10^{10\pm3}$ $s^{-1}$. Deviation in the parameters for different measurement of different samples is about 30%.

Thermal dependences for both centres due to noble ions are shown in parts (b, c) of Fig.3. For them there is no drop in intensity near 200 K, as it happens in the sample without noble ions, as shown in the part (a), Fig.3. For the case of $Cu^+$ center a good correspondence between thermal dependences of time constant and intensity is observed. For the case of $Ag^+$ center there is a bump in the dependence of intensity, which has a counterpart in the thermal dependence of the time constant. The $Ag^+$ luminescence time constant decreases from 27 μs to 20 μs in the range of 100 K -300 K, which is accompanied by an increase in the intensity. That could be related to change of transition probability but only partly, because decrease in time constant is not so pronounced. Perhaps, increase in intensity could be related to activation of some defects with silver silent in luminescence.

Thermal quenching is started at 450 – 500 K for both noble ions centers, Fig.3 b and c parts. The quenching parameters for noble ions centers are: (Ag) $E(I) \sim 0.65 \pm 0.05$ eV; $E(\tau)=0.75 \pm 0.05$ eV; $f(I)=3 \cdot 10^{9 \pm 3}$ s$^{-1}$; $f(\tau)=2.4 \cdot 10^{13 \pm 3}$ s$^{-1}$; (Cu) $E(I) = 0.65 \pm 0.05$ eV, $E(\tau) = 0.63 \pm 0.05$ eV, $f = 2.4 \cdot 10^{12 \pm 3}$ s$^{-1}$.

At temperature below 50 K, for all the studied centers there are observed time constant changes, which are not accompanied by changes in intensity, see Fig.3. In Fig.4 decay kinetics curves for different temperatures are presented. In all cases long components appeared at low temperatures 3 K and 5 K. In Al – Me$^+$ case (Fig.3 a) a fast component could be observed at 3 K and 5 K. This behavior is typical for the splitting of the triplet state in zero magnetic field. Thus, the radiating levels of the center of noble ions could be triplet. The excited states of Cu$^+$ and Ag$^+$ contain $^3D$ term which, after Hund rule, is certainly lowest [10]. The alkali related blue luminescence of a quartz crystal is also due to a triplet-singlet transition, [2].

*Thermally and optically stimulated luminescence in X-ray irradiated α-quartz*

The luminescence spectra of X-ray excited α-quartz samples contain the same bands as in the case of photoluminescence. In the quartz samples doped with silver, the main band is at 4.75 eV. In copper doped samples, the main band is at 3.4 eV, and in natural α-quartz samples (smoky, morion, etc.), the main band is at 3 eV. Under continuous x-ray excitation, the intensity of these bands varies with time, as shown in Fig.5. As a rule, such behavior corresponds to the complex processes of interaction of the definite luminescence centers with an electron-hole cloud undergoing charge capture and subsequent recombination.

Fig. 6 shows the thermally stimulated luminescence of the studied samples in the high-temperature range. There are several TSL peaks with main luminescence bands, coinciding with photoluminescence bands of corresponding samples. The position of TSL peaks is different for different samples, showing on different traps. At an irradiation temperature (293 K), a afterglow curves are observed, Fig. 6, inserts. The kinetics curve does not straighten in a double logarithmic scale. The decay kinetics law is complicated. At the end of the decay, it can be approximated by the hyperbolic law $\sim t^{-n}$, where n is varying from 0.25 to 0.65. Such low value of exponent is not typical for monomolecular and bimolecular recombination kinetics (see, for example, [10]).

In the quartz crystal doped with copper the Cu$^0$ centers are very efficiently produced by X-ray irradiation at a low temperature. This is proved by the electron spin resonance (ESR) method, [11], where the ESR signal of Cu$^0$ and induced absorption band at 4.2 eV bleached in similar manner with photons ~4 eV. The corresponding spectra of induced absorption are described in literature [12]. Fig. 7 shows the thermal annealing of the induced absorption corresponding to the center of Cu$^0$. The insert shows the TSL curve during annealing for low temperature range. The TSL peak is located at 244 K, as it was published earlier in [12]. It can be seen that during annealing a decrease in the absorption intensity takes place as well as redistribution between induced absorption bands occurs with the appearance of a new band at 5 eV, Fig.8. This absorption band of 5 eV appears in a narrow temperature range on the high-temperature side of the TSL peak. In Fig.8 the intensity of the band at 4.2 eV growth up at 110 K accompanied with drop in intensity of the band at 3.2 eV. Following heating reduces the intensity of the band at 4.2 eV accompanied with increase of intensity of the band at 3.2 eV. In the range of 240 K, where both bands at 4.2 eV and 3.2 eV drop down, the band at 5 eV strongly increases and also decreased to 250 K. Having in mind that the band at 4.2 eV belongs to $Cu^0$ center [11], then increase in the range of 110 K can be due to increase of correspondent center concentration and the source is the center responsible for 3.2 eV band. Then the band at 3.2 eV is also $Cu^0$ perhaps in another structural position. Disintegration of both centers with the band at 3.2 eV and 4.2 eV provide band at 5 eV, which then is also due to $Cu^0$ center in third position. Therefore, all three bands are attributed to the center of $Cu^0$ located in different interstitial positions. A thermally activated exchange occurs between these positions. The absorption bands can be bleached by irradiation with photons 3; 4; 5 eV. Photo bleaching with photon 3 eV energy is accompanied by optically stimulated luminescence, which corresponds to the $Cu^+$ center with a band at 3.4 eV (360 nm), Fig.9. Therefore, it can be concluded that the thermally stimulated release of $Cu^0$ from different positions and motions is the basis for the creation of the TSL peak. $Cu^0$ reaches a hole on the aluminum-oxygen defect, recombines with a hole, and converts into $Cu^+$ in the excited state.

For the case of a silver-doped natural quartz crystal, a high-temperature TSL contains more peaks than the copper doped quartz, Fig.6. The luminescence band in the case of silver is 4.75 eV (280 nm). It can be noted that the first TSL peak at 375 K practically coincides with the TSL peak of copper center. However, in the case of silver, the intense peaks are observed also at 455 K and 520 K. Afterglow recorded after X-ray irradiation at 293 K (Inset of Fig.6), exhibits a hyperbolic decay, also approximated by ~ $t^{-0.5}$, which is not directly interpreted in the literature as in the case of copper. Irradiation at 80 K provides a TSL peak at 180 K. Quartz crystal doped with silver, after irradiation at 80 K displays an ESR signal of the $Ag^0$ centers [13]. Radiation induced silver center $Ag^0$ in quartz, in addition to the ESR signal, provides luminescence center with a band at 2.8 eV [5, 14]. It was noted that heating of the irradiated quartz crystal with a silver dopant changes position of the induced luminescence band from 2.8 eV to 2.3 eV [5, 14], Fig. 10. The TSL curves corresponding to the different bands are shown in Fig. 10, insert. It can be seen that the band at 2.3 eV appears on the high-temperature side of the main TSL peak. For induced luminescence center with a band at 2.3 eV a corresponding ESR signal was not detected, [14] contrary to the induced luminescence center with the band at 2.8 eV. On the other hand, excited by photon in the intra-center process, the decay of the induced band at 2.3 eV is slow (~ 200 μs [15, 16, and 17]). This was explained by the forbidden, possibly triplet-singlet transitions. the observed effect was explained by thermally stimulated movement of $Ag^0$ and collecting near $Ag^+$ with creating of a cluster $Ag_3^+$ [15, 16, 17]. It should be ESR silent [18], diamagnetic, and can explain the existence of triplet and singlet excited states responsible for the fast and slow components of the irradiated luminescence center in quartz [16, 17].

The $Ag^0$ center, which provides the luminescence band at 2.8 eV, can be destroyed by light, and accordingly the induced absorption bands can be photobleached. During such photobleaching, we observe optically stimulated luminescence, which spectrum is similar to presented in Fig.10 for 180 K. Therefore the luminescence of $Ag^0$ and $Ag^+$ are observed in OSL. The OSL kinetics curve is shown in Fig.11. In the end of decay this kinetics can be approximate with power law $t^{-1.64}$ which is close to $t^{-2}$ or kinetics of bimolecular recombination.

The center responsible for 2.3 eV attributed to the $Ag_3^+$ cluster cannot be destroyed by light, it can be thermally destroyed only at very high temperature. In extreme cases, some samples after irradiation become yellow. This is explained by creation of colloidal silver. Thus, in the cases of of quartz crystal doped with copper and silver, the effect of photo thermo stimulated migration of radiation defects $Cu^0$ and $Ag^0$ was observed and the recombination luminescence is observed as result of such migration.

In a natural quartz crystal before the introduction of copper or silver we can observe the afterglow and TSL after ceasing of X-ray irradiation, Fig. 12. Irradiation provides induced absorption, which can be optically destroyed with observation of OSL. The OSL spectrum coincides with the photoluminescence spectrum of this crystal, Fig. 12. It is possible assume migration of alkali atoms in this process, as in the case of samples with copper and silver, however those atoms are not yet identified.

Discussion

Two types of impurity luminescence centers in crystalline quartz were studied in a wide temperature range of 3-700 K. One of the centers present in natural crystals gives a blue luminescence at 3eV, it was previously studied in [2, 3] in the temperature range of 3-300 K. The nature of luminescence center is ascribed to a complex defect containing aluminum ions and alkali ions. Second type center was obtained by modification first type center by substitution of alkali ions for noble ions - copper or silver. The centers of noble ions were also studied earlier in the same temperature range 4-300 K [4, 5]. It is remarkable that the subsequent reverse substitution of noble ions for alkali ions restores the initial luminescence type one. That means that during both exchanges the initial defect configuration is the same for both kinds of centers. Now the attention was paid to the study of quenching of luminescence. Quenching almost completely occurs at 700 K for centers with alkali ions from one hand and noble ions from the other. Different values of the activation energy of quenching are obtained. For the alkali related center, it is about ~0.2 eV, and for the center of noble ions it is about ~0.7 eV. Possible reasons for this may be related to the differences between alkali ions and noble ions. Let us consider the possible models of the luminescence center. The structure of the centers was proposed as Al, O defect surrounded in the first case by alkali ions and in the second case with noble ions [3, 4, 5].

Actually the real structure of defect with aluminum is not clear, [3], where it was presumed that [AlO$_4$] possibly could not participate in studied luminescence center. However, this question needs special investigation. For alkali ions, electronic transitions from the filled shells have a high energy of about 50 eV, that means that electronic transitions of the luminescence center can have a charge transfer type when an electron from the p-states of oxygen passes into the s state of alkali ions or the complex Al, O$^-$, Me$^+$ + ℏν => Al, O$^o$, Me$^0$. Thus, the situation is very similar to the properties of the so-called L-center of sodium-silicate glasses, [19]. Aluminum itself does not create a center of luminescence in silicon dioxide [6, 7]. There is no quartz crystal samples available doped with aluminum only, that is why this proposition cannot be verified directly, but there is a study of silica glass doped with aluminum without alkali ions [7]. In this case, aluminum stimulates only the creation of an oxygen-deficient center of luminescence [7]. In silica glass, doped together with aluminum and alkali ions, the luminescence center is, in some way, analogous to the one under discussion [6]. In the silica glass with aluminum and alkali ions, the substitution of alkali ions for noble ions also leads to the creation of a luminescence center associated with noble ions as well.

For Al, O$^-$, Cu$^+$ (Ag$^+$) defect in a quartz crystal the properties of the luminescence center are determined mainly by intra-ionic transitions in noble ions (d10⇔d9s), as was interpreted for the luminescence center of noble ions in alkali halides, [20, 21, 22]. The difference in the properties between the silica glass and the quartz crystal lies in the different geometric structure of the corresponding center. In silica glass without aluminum, the center is created on the basis of non-bridging oxygen, whereas in a crystal a defect of the non-bridging oxygen is very rare. In crystal quartz with aluminum, the center is created without the participation of a defect of non-bridging oxygen. The one valence ion is in some interstitial position near the defect created by the presence of aluminum. Study of smoky color in quartz shows [23] that a hole is trapped on AlO$_4$ tetrahedron. Alkali ion traps an electron and alkali atom moves away similarly to the case of Cu$^0$ and Ag$^0$ described in previous section.

The question arises as to why these centers have the final thermal quenching at a relatively high temperature. This makes it possible to use natural crystalline quartz in radiation measurements and dosimetry (see e.g. [24]). First of all, the blue center of luminescence in natural quartz has the strong thermal quenching, beginning at 180 K, (Fig.3) greatly decreasing the luminescence intensity by 300 K. At the same time, similar quenching was observed for the luminescence of a self-trapped exciton (STE) [2], which has a similar activation energy for thermal quenching (0.2 - 0.3 eV). In addition, the center of blue luminescence has singularities at 5-20 K in the decay kinetics analogous to that for STE, which are explained as triplet splitting in a zero magnetic field. This allows the preposition of generation of STE-like structure near the aluminum defect, similar to the generation of STE near the germanium impurity in a quartz crystal [2]. The model of STE is based on assumption that silicon-oxygen bond is broken (weakened). The relaxation leads to creation of a bond between appeared transient non-bridging oxygen and bridging oxygen of opposite site of C or x,y channel of the α-quartz structure. The thermal stability of STE is related to the strength of the non-bridging-oxygen-bridging-oxygen bond of the STE base in silicon dioxide [25]. This oxygen-non-bridging-oxygen bond appears in the STE with a hole on oxygen and an electron on silicon.

During excitation of the blue luminescence center an electron appears on an alkali due to charge transfer transitions from the oxygen states p. Perhaps, oxygen with a hole creates a bond with nearest oxygen. Thermal stability of this bond determines quenching at 200 K, as in the case of STE. For an excited center with a noble ion oxygen has not have a hole and therefore lack of noble ions luminescence quenching at 200 K. Thus, for the alkali related center of blue luminescence, two factors are assumed: the strength of oxygen-oxygen bond at low temperatures and presence of the alkali atom in the interstitial position at high temperatures. The experimental data (Fig. 4) show that the acceleration of the decay kinetics with increasing temperature due to radiative transition probability increasing with heating. The intensity of PL depends on the temperature with a delay relative to the time constant. This can be seen from the difference in the values of the frequency coefficient, determined from $\tau$ (T) and I (T) for the blue luminescence. The strength of oxygen-oxygen bond determines the main thermal quenching at a temperature of about 200 K, but some centers can emit at high temperatures because of the stability of the alkali atom. Oxygen-oxygen bond, of course, is destroyed, but the hole remains on oxygen belonging to aluminum defect.

The process of thermal quenching in different temperature regions can be explained by different intersection of adiabatic surfaces of ground and excited states. At low temperature the structure of aluminum-alkali excited center comprises a hole on the oxygen-oxygen bond and an electron on an alkali. We can assume that symmetry of the excited and ground states are different and in such situation crossing of the excited state with the ground state is possible. And high temperatures, when oxygen-oxygen bond is broken, the center structure is changed. The symmetry of the ground and excited states could be the same and crossing of the excited and ground states are avoided. Therefore the center is completely thermally quenched at high temperatures.

In the case of a center with noble ions, the intra-ionic transitions take place. It is assumed that an oxygen-oxygen bond is not created because there is excited only the noble ion, and therefore there is no luminescence similar to STE as well as the corresponding thermal quenching at 180 K. There is an intra-ionic quenching. Excited ions perform thermal deactivation only at high temperatures.

The model of the radiation-induced process is more understandable for the case of crystals doped with copper and silver. Under ionizing irradiation, a hole from an electron-hole cloud is obviously captured by oxygen connected to aluminum and an electron is captured on the nearest one valence ion. A positive compensator is no longer needed, and the created atom performs forced diffusion from the initial position. Thus, $Cu^0$ and $Ag^0$, as well as not yet detected, $Me^0$ (Me = Li or Na or K, etc.) appeared and stabilized in some interstitial wells. These atoms can change the well in a thermally stimulated process (Fig. 6, 7). Interstitial atom migration provides the TSL. In the case of silver, the migration of atoms can provide even a larger silver clusters.

Generation of such clusters takes place also for copper or alkali metals. Such clustering may be responsible for TSL peaks at high temperatures, Fig.6.

Conclusions

The luminescence center in α-quartz crystal consisting of aluminum, oxygen and one of valence compensating ions (alkali ions or noble ions of $Cu^+$ or $Ag^+$) exhibits thermal quenching at 700 K. In the case of noble ions, intra-ionic transitions determine the properties of the process associated with luminescence and quenching. In contrast, in alkali ions intra ionic excitation should happens at too high energy, that is why only the charge transfer transition takes place in the this center. A hole of the excited center remains on an oxygen, and an electron stays on an alkali. The center is similar to STE in quartz with the creation of an oxygen-oxygen bond, when a hole is in the p-state, and an electron is on an alkali. Therefore, the quenching processes are different for alkali and noble ions. The center with alkali ions exhibits complex quenching, modeled by two processes. One of them is a low-temperature (200 K) quenching attributed to a process analogous to quenching of STE luminescence, which is due to the oxygen-oxygen bond strength. In this process, the most of the excited centers are extinguished. However, a less probable process is possible. The electron remains on the alkali, which stimulates the stability of the hole in oxygen. The difference could be explained by different crossing of the excited and ground state adiabatic surfaces in two regions of thermal quenching. At low temperature there is crossing of these surfaces and at high temperatures the crossing are avoided. The thermal stability of an alkali atom provides a less intense luminescence at a high temperature with the same spectral parameter as the low-temperature luminescence.

Ionizing irradiation provides effective processes of radiation-stimulated diffusion of ion compensators. A positive ion compensator is an effective trap for an electron. Such an atom is localized in different wells, providing a spectrum of radiation-induced defect. Photo-thermal stimulation provokes diffusion of these interstitials and recombination with a trapped hole on aluminum containing a defect in samples of natural quartz crystals that generate the TSL and OSL signals.

ACKNOWLEDGMENTS
This work was supported by the Latvian Science Council Grant No lzp-2018/1-0289 .

Figures caption

Fig.1

Photoluminescence (PL) spectra, time constant spectra (closed squares) under ArF laser excitation, inset – decay kinetics. Photoluminescence excitation (PLE) (80 K) and absorption (293 K) spectra of synthetic (short dashes) and natural (lines) Al doped quartz.

Fig.2

PL spectra of silver and copper doped (lines, 293 K) and untreated natural quartz (dashes dots, different temperatures) samples excited with pulses of ArF laser.

Fig.3

Temperature dependences of PL intensity (lines) and time constant (points) of natural quartz morion (dashes) and smoky (lines) (part a).
Temperature dependences of PL intensity (lines) and decay time constant (points) of copper (360 nm, part b, $Cu^+$ I(T) τ(T)) and silver (260 nm, part c, labeled $Ag^+$ I(T) τ(T)) doped natural quartz. Excitation –ArF excimer laser (193 nm).

Fig.4

PL decay kinetics curves measured under excitation with the ArF excimer laser (193 nm) in smoky quartz before electrolysis (a, time scale in ms) and after electrolysis of copper (b, time scale in µs) and silver (c, time scale in µs). Registration is done with the digital oscilloscope. The cures for 3 K and 5 K are measured under excitation with spark discharge light through the vacuum monochromator, selecting photon energy 6.4 eV. Registration is done with the multichannel analyzer.

Fig.5

Kinetics of luminescence of natural α-quartz under continuous x-ray excitation.

Fig.6

TSL and afterglow (insets) of natural α-quartz (Part (a) labeled Al – Me, the band at 400 nm); of Cu doped (Part (b) labeled Al – Cu, the band at 360 nm); of Ag doped (Part (c) labeled Al – Ag, the band at 260 nm) of natural α-quartz after X-ray irradiation. Insets also present power law fits of the afterglow curves.

Fig.7

Thermal annealing of the induced optical absorption of X-ray irradiated Cu doped natural α-quartz. Insert – TSL peak at 244 K selecting the luminescence band 360 nm.

Fig.8

Thermal dependences of the absorption bands induced by X-ray irradiation in Cu doped natural α-quartz.

Fig.9

OSL spectrum of x-ray irradiated Cu-doped a-quartz, T=80 K. The spectrum measured with resolution ~0.05 eV using Hg discharge light source with the 405 nm selective optical filter.

Fig.10
Luminescence spectra of Ag-doped a-quartz observed under X-ray excitation at different temperatures related to thermally stimulated luminescence. Insert – TSL curves for different luminescence bands.

Fig.11
OSL of x-ray irradiated at 80 K silver doped α-quartz. Stimulation done at 3.5 eV.

Fig.12
OSL (stimulation at 4 eV) and induced absorption spectra of morion α-quartz T=80 K.

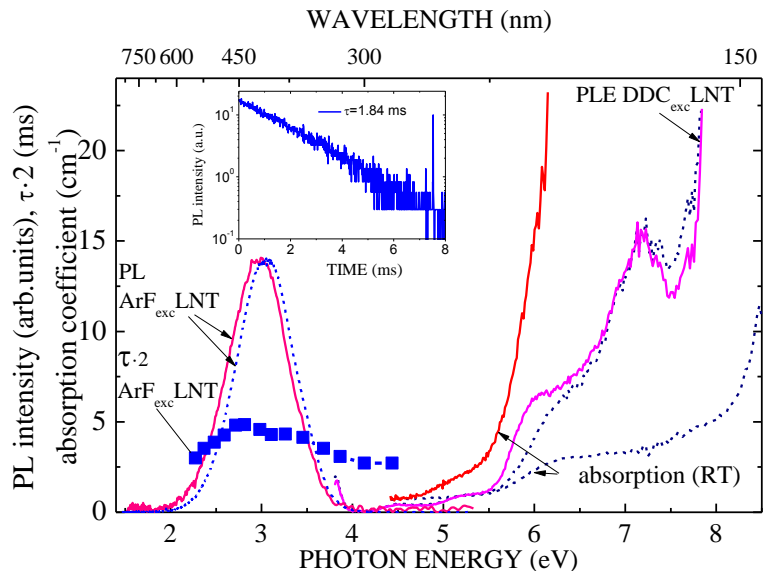

Fig.1

Photoluminescence (PL) spectra, time constant spectra (closed squares) under ArF laser excitation, inset − decay kinetics. Photoluminescence excitation (PLE) (80 K) and absorption (293 K) spectra of synthetic (short dashes) and natural (lines) Al doped quartz.

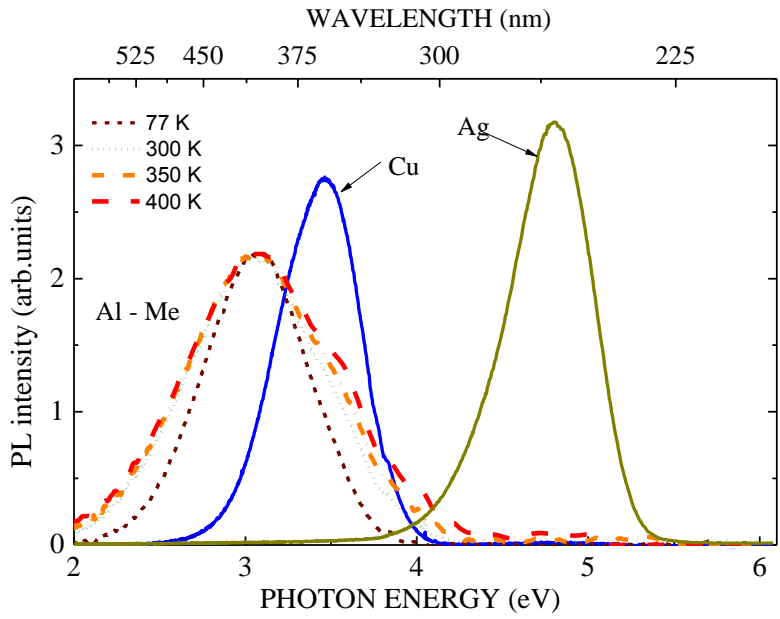

Fig.2
PL spectra of silver and copper doped (lines, 293 K) and untreated natural quartz (dashes dots, different temperatures) samples excited with pulses of ArF laser.

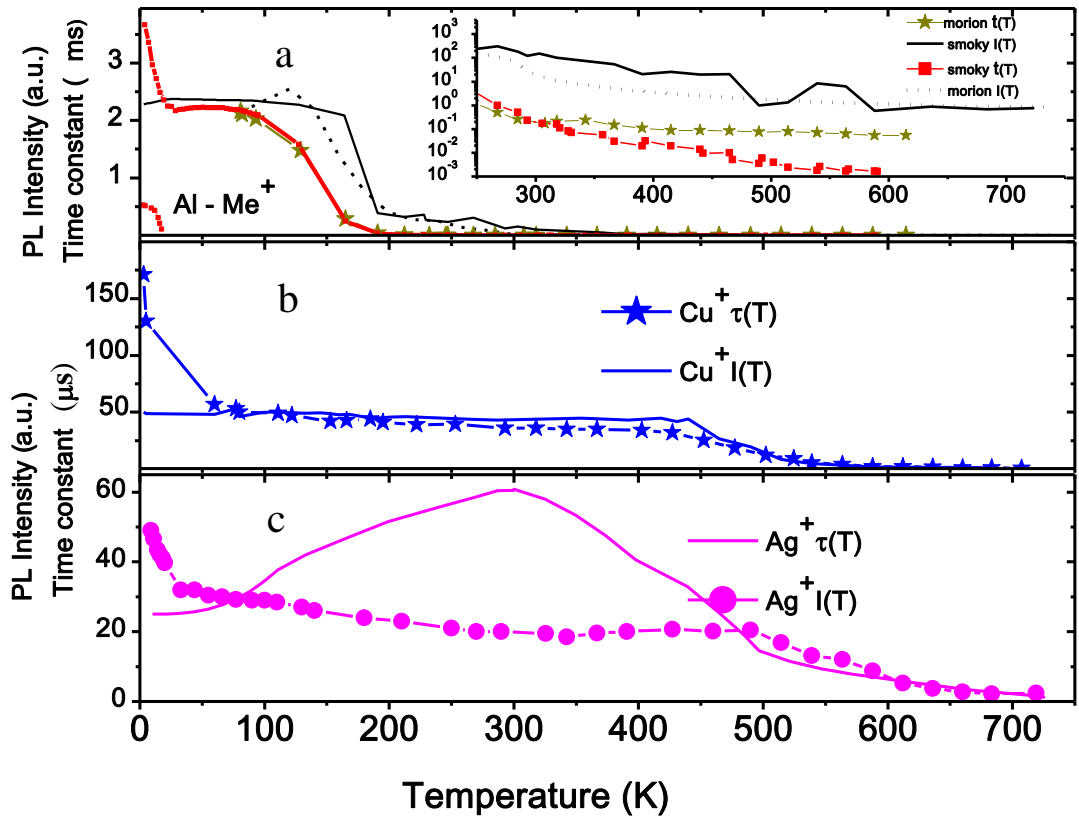

Fig.3
Temperature dependences of PL intensity (lines) and time constant (points) of natural quartz morion (dashes) and smoky (lines) (part a).

Temperature dependences of PL intensity (lines) and decay time constant (points) of copper (360 nm, part b, $Cu^+$ I(T) τ(T)) and silver (260 nm, part c, labeled $Ag^+$ I(T) τ(T)) doped natural quartz. Excitation –ArF excimer laser (193 nm).

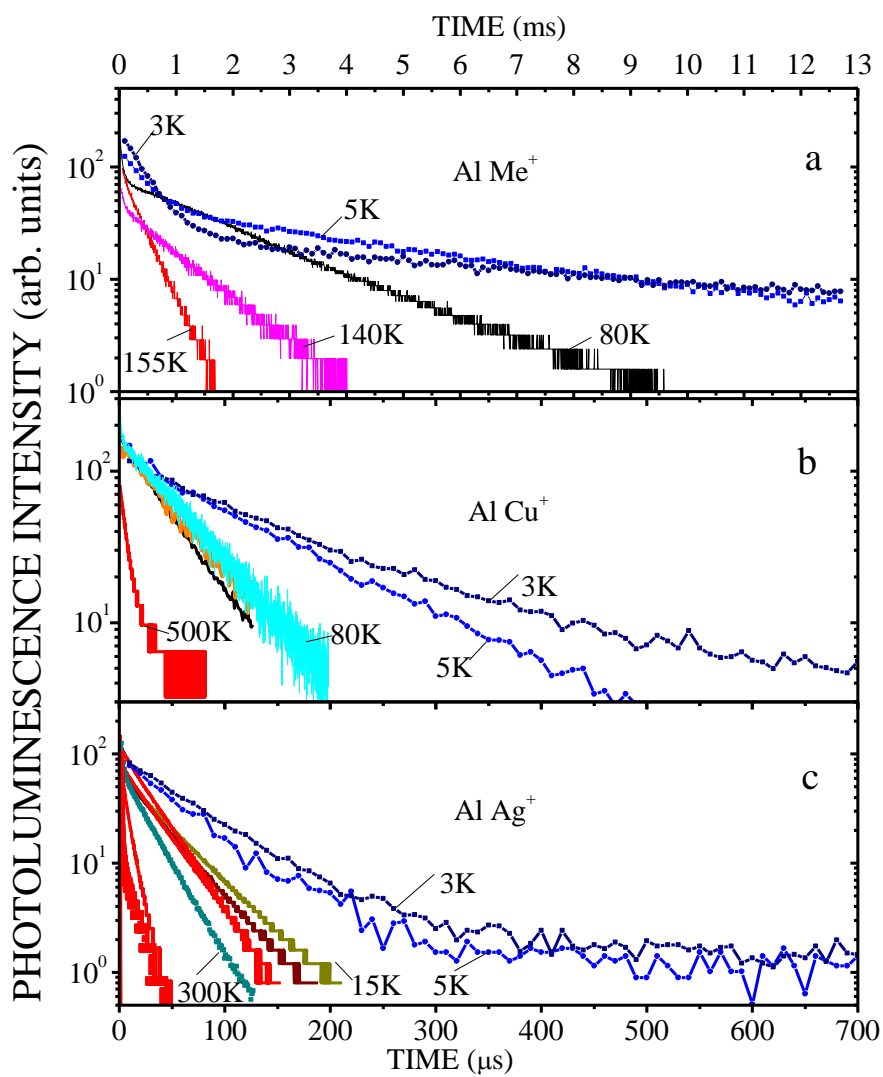

Fig.4
PL decay kinetics curves measured under excitation with the ArF excimer laser (193 nm) in smoky quartz before electrolysis (a, time scale in ms) and after electrolysis of copper (b, time scale in µs) and silver (c, time scale in µs). Registration is done with the digital oscilloscope. The cures for 3 K and 5 K are measured under excitation with spark discharge light through the vacuum monochromator, selecting photon energy 6.4 eV. Registration is done with the multichannel analyzer.

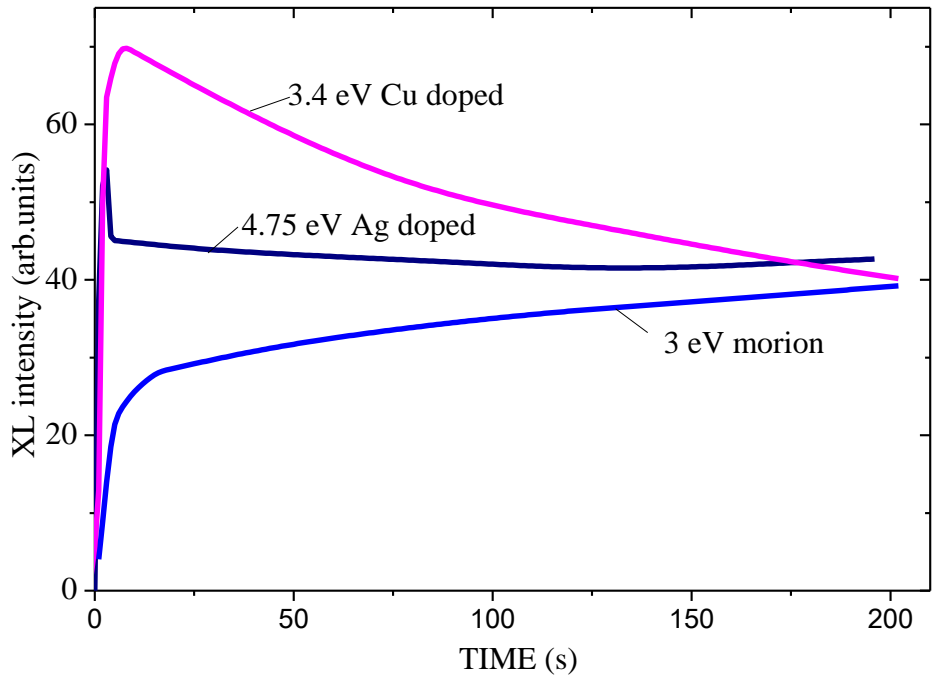

Fig.5
Kinetics of luminescence of natural a-quartz under continuous x-ray excitation.

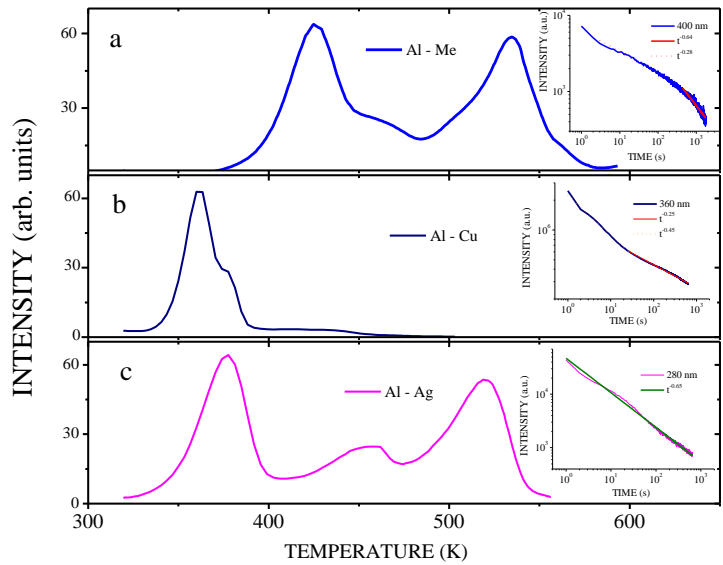

Fig.6

TSL and afterglow (insets) of natural α-quartz (Part (a) labeled Al – Me, the band at 400 nm); of Cu doped (Part (b) labeled Al – Cu, the band at 360 nm); of Ag doped (Part (c) labeled Al – Ag, the band at 260 nm) of natural α-quartz after X-ray irradiation. Insets also present power law fits of the afterglow curves.

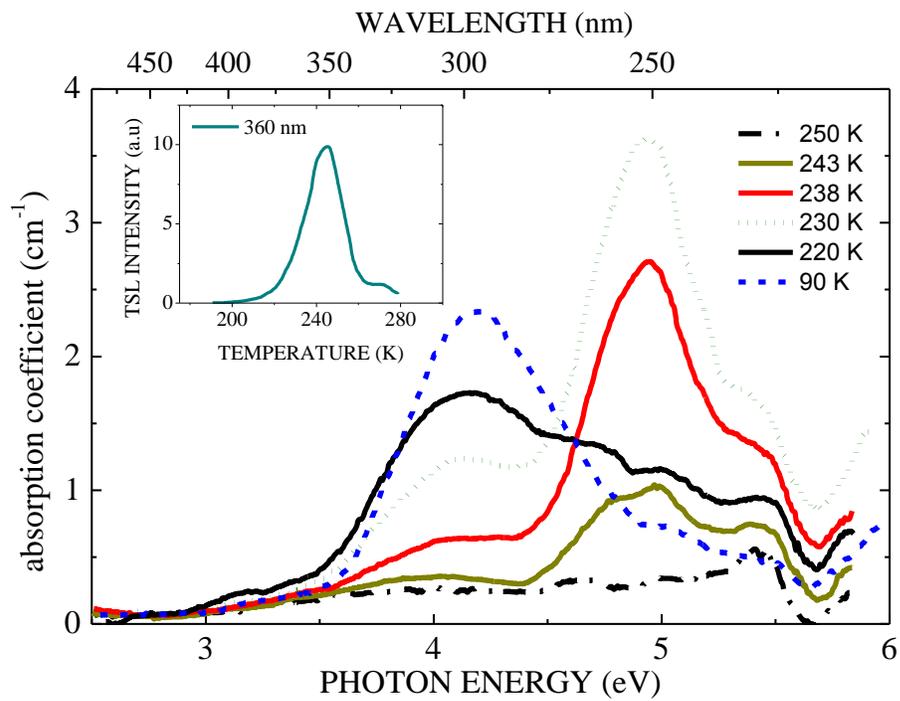

Fig.7
Thermal annealing of the induced optical absorption of X-ray irradiated Cu doped natural α-quartz. Insert – TSL peak at 244 K selecting the luminescence band 360 nm.

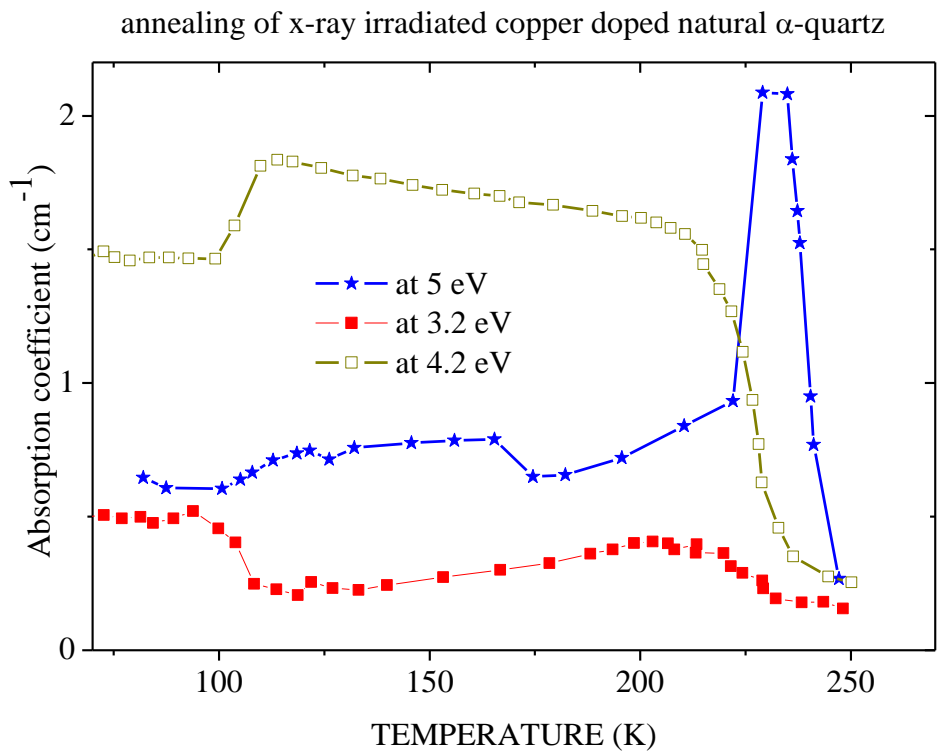

Fig.8
Thermal dependences of the absorption bands induced by X-ray irradiation in Cu doped natural α-quartz.

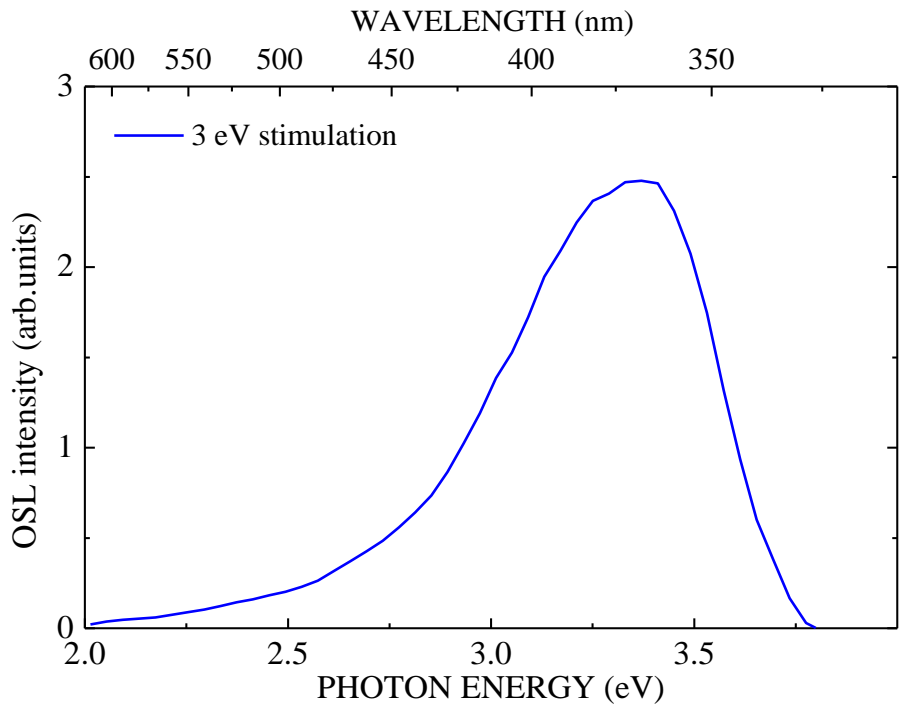

Fig.9
OSL spectrum of x-ray irradiated Cu-doped a-quartz, T=80 K. The spectrum measured with resolution ~0.05 eV using Hg discharge light source with the 405 nm selective optical filter.

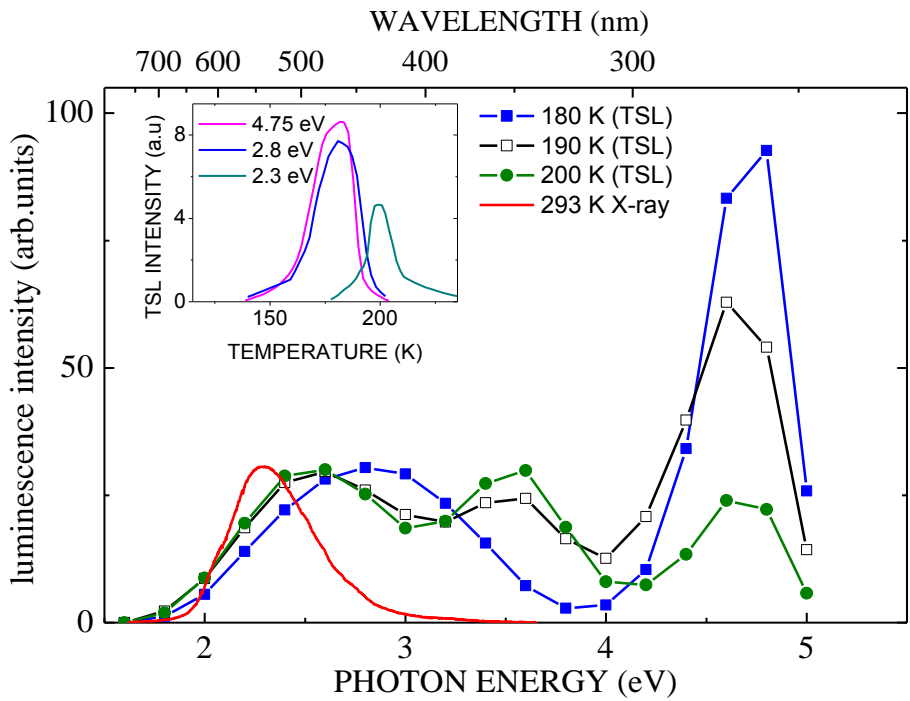

Fig.10
Luminescence spectra of Ag-doped a-quartz observed under X-ray excitation at different temperatures related to thermally stimulated luminescence. Insert – TSL curves for different luminescence bands.

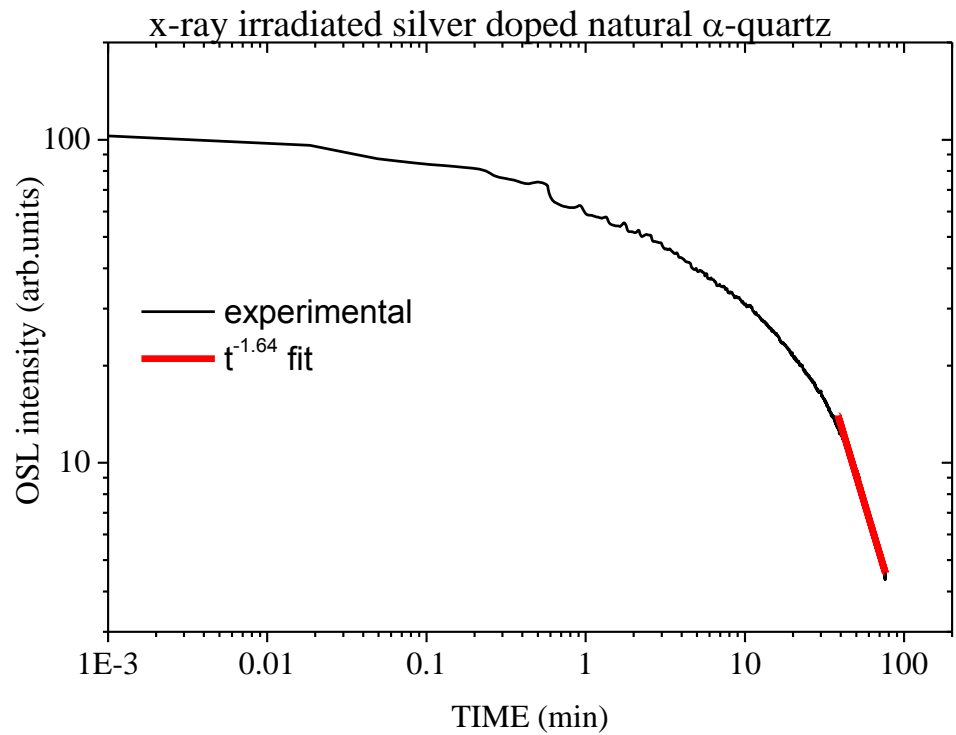

Fig.11
OSL of x-ray irradiated at 80 K silver doped α-quartz. Stimulation done at 3.5 eV.

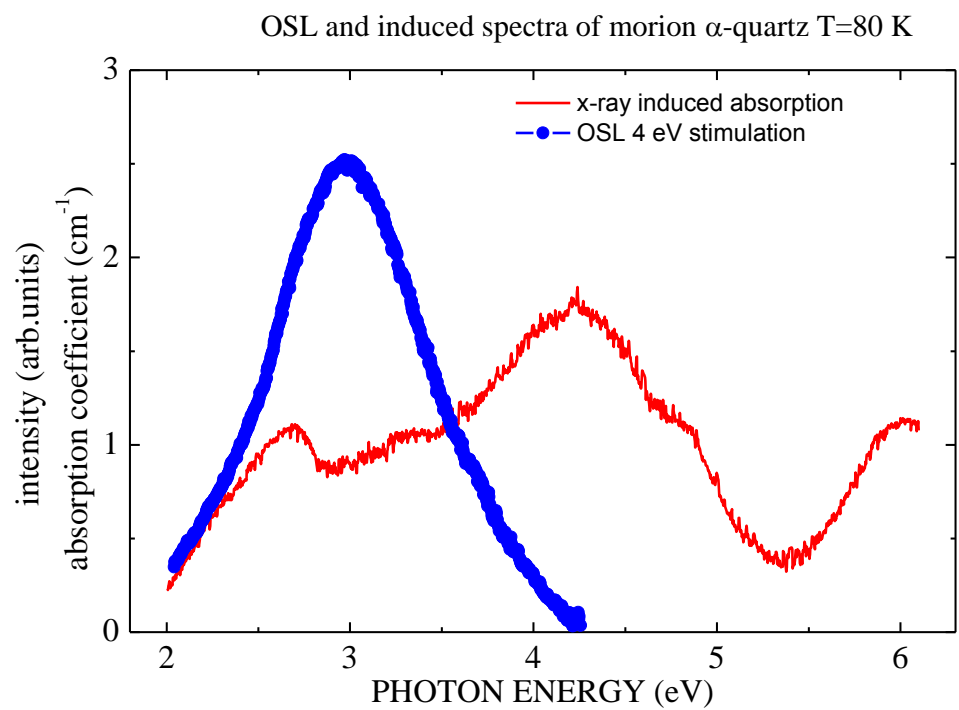

Fig.12
OSL (stimulation at 4 eV) and induced absorption spectra of morion α-quartz T=80 K.